\documentclass{aastex}

\begin{document}

\title{Time-series Spectroscopy of Pulsating \lowercase{sd}B Stars:
  PG1605+072\footnote{Based on observations made with the Danish
  1.54\,m telescope at ESO, La Silla, Chile, from the South African
  Astronomical Observatory (SAAO), and from Mt Stromlo Observatory,
  Australia.}}

\author{S. J. O'Toole and T. R. Bedding}
\affil{School of Physics, University of Sydney, NSW 2006, Australia}

\and

\author{H. Kjeldsen}
\affil{Theoretical Astrophysics Center, Aarhus University, DK-8000,
  Aarhus C, Denmark}

\and

\author{T. C. Teixeira}
\affil{The MONS Project Team, Institute of Physics and Astronomy,
  Aarhus University, DK-8000 Aarhus C, Denmark}

\and

\author{G. Roberts, F. van Wyk and D. Kilkenny}
\affil{South African Astronomical Observatory, PO Box 9, Observatory
  7935, South Africa}

\and

\author{N. D'Cruz and I. K. Baldry}
\affil{School of Physics, University of Sydney, NSW 2006, Australia}

\newpage

\begin{abstract}

We report the detection of velocity variations in the pulsating
sdB star, PG 1605+072. Oscillations are detected at the same frequencies
found from photometry and have amplitudes of up to
14\,km\,s$^{-1}$ for H$\beta$. The strongest oscillation found in previous
photometric observations is not evident in our spectroscopy or
photometry, and may be absent due to beating of closely spaced
modes. Phase differences between spectroscopy and B magnitude
photometry imply that maximum brightness occurs not long after maximum
radius. We have also found evidence of variation in the observed
amplitudes of five Balmer lines, with a decrease in amplitude of the
strongest mode blueward from H$\beta$. This effect is not expected and
a longer time-series will be needed to clarify it.

\end{abstract}

\section{Introduction}

The discovery of a class of pulsating hot subdwarfs (sdBs;
\citealt{ECpaperI}) has created an interest in asteroseismology of
evolved stars. Subdwarfs are believed to lie on the Extreme
Horizontal Branch \citep{SBK94}, although their formation and
evolution is still uncertain. Pulsations in sdBs were predicted
theoretically by \citet{CFB96} just prior to their discovery, and
about 20 pulsators have been discovered to date using time-series
photometry (see \citealt{ECpaperXIII} for a review). The stars (also
known as EC14026 stars, after the prototype) are believed to pulsate
in $p$ modes and typically have periods 100--200\,s and
semi-amplitudes $<$10\,mmag. Models indicate an opacity bump associated
with iron ionisation may be the driving mechanism of the pulsations
\citep{CFB96,CFB97a}, although recent studies by
\citet{ECpaperXIII} find no well-defined instability strip in the
log\,$g-T_{\mathrm{eff}}$ plane.

Combining time-series photometry with time-series spectroscopy
should help to give a better understanding of the pulsations in these
stars. We have shown the feasibility of time-series
spectroscopy with observations of PG1605+072 on a 1.5\,m telescope
over 2 hrs \citep{OBK00}. \citet{J+P00} made
similar observations of KPD2109+4401 (5\,hrs) and PB8783 (6\,hrs)
using the 4.2m William Herschel Telescope. In all three cases,
velocity variations were detected at the published photometric
frequencies. However, sdB stars are multimode pulsators and longer
time series are needed to resolve the individual modes and measure
their amplitudes. We present here the first extended time-series
spectroscopy of a pulsating sdB star.

Our target was PG1605+072 which has the most extreme properties of all
the pulsating sdBs detected so far, with high amplitudes (up to
$\sim$25\,mmag) and long periods ($\sim$500\,s). Because of its
low gravity (log\,$g$\,$\sim$5.25), this star is believed to have evolved
off the Horizontal Branch and may oscillate in $g$, as well as $p$
modes \citep{ECpaperX}. The power spectrum is complex, with up to 55
indentifiable frequencies. PG1605+072 displays considerable rotation
($v$sin$i=39$\,km\,s$^{-1}$), possibly leading to unequally spaced
multiplet components \citep{HRW99}. Unlike several of the pulsating
sdBs, multicolour photometry gives no indication of a cool companion
\citep{ECpaperVII}.

\section{Observations}

We obtained medium-resolution spectra of PG1605+072
using the DFOSC spectrograph on the Danish 1.54\,m
telescope at La Silla, Chile and the coud\'e spectrograph (A grating)
on the 74-inch (1.88\,m) telescope at Mt Stromlo, Australia. The
observations were made on 7 nights over an 11-day period in July and
August, 1999 (see Table \ref{spec}). To supplement the spectroscopy,
we obtained time-series photometry at the Sutherland site of the South
African Astronomical Observatory (see Table \ref{phot}).

\placetable{spec}
\placetable{phot}

The La Silla data consisted of single-order spectra projected onto a 2K
LORAL CCD - pixel-binning (to reduce R.O.N.) and windowing (to reduce
readout time) gave 66 x 500 pixel spectra with a total wavelength
range of 3700--5000\,\AA\ and a dispersion of
1.65\,\AA\,pixel$^{-1}$. The resolution was 2.8\,\AA, set by a slit
width of 1.5 arcsec. The exposure time was 46\,s, with a dead time of
about 15\,s. The average number of photons per \AA\ in each spectrum
was about 2200.

The Mt Stromlo data consisted of single-order spectra projected onto a
SITe 2Kx4K CCD, with a wavelength range of 3800--5000\,\AA. The slit
width was $\sim$4 arcsec and the dispersion was
0.60\,\AA\,pixels$^{-1}$, leading to a resolution similar to that of
the La Silla data. The exposure time was 50\,s with a dead time of
around 25\,s.

The photometry from Sutherland was taken using the Modular Photometer with a
Johnson $B$ filter and exposure times of 20\,s. The GaAs
photomultiplier tube in this photometer has a good red sensitivity, and
using the blue filter causes the passband to resemble that of a
blue-sensitive photomultiplier tube \citep{ECpaperVI}.

\section{Reductions}

Standard methods for bias subtraction, flat-fielding
and background scattered-light subtraction were used.
Non-linearities on the CCD image were only important in
the Stromlo data, and were corrected immediately after bias
subtraction \citep[see]{BBV98a}. One-dimensional spectra were extracted
using the Optimal Extraction method \citep{OptEx86}, a variance
weighting system for CCD columns. A 3rd order polynomial was fitted to
the continuum level in both data sets, and the spectra were normalised
to a continuum value of unity. 

The spectrum of PG1605+072 is dominated
by Balmer lines - see e.g. Figure 9 of \citet{ECpaperVII}.
A cross-correlation technique was used to determine the Doppler shift
for five Balmer lines (H$\beta$, H$\gamma$, H$\delta$, H$\epsilon$ and
H8), relative to a template spectrum (the average of 20 high-quality
spectra).
The spectra and template were prepared in the manner described by
\citet{BBV98a}. It should be noted that the time-series for
H$\gamma$ does not contain any Mt Stromlo data, because that line was
contaminated by a bad CCD column.

Several jumps were found in each time-series,
particularly in the La Silla data, occurring every 30-40
observations, corresponding to breaks where the He+Ne calibration
spectra were taken. We decided to process each of these groups
separately with a template spectrum created for each group. In
removing the jumps, we have effectively high-pass filtered the
dataset, also removing any instrumental drifts present. 

\begin{figure}[h]
\plotone{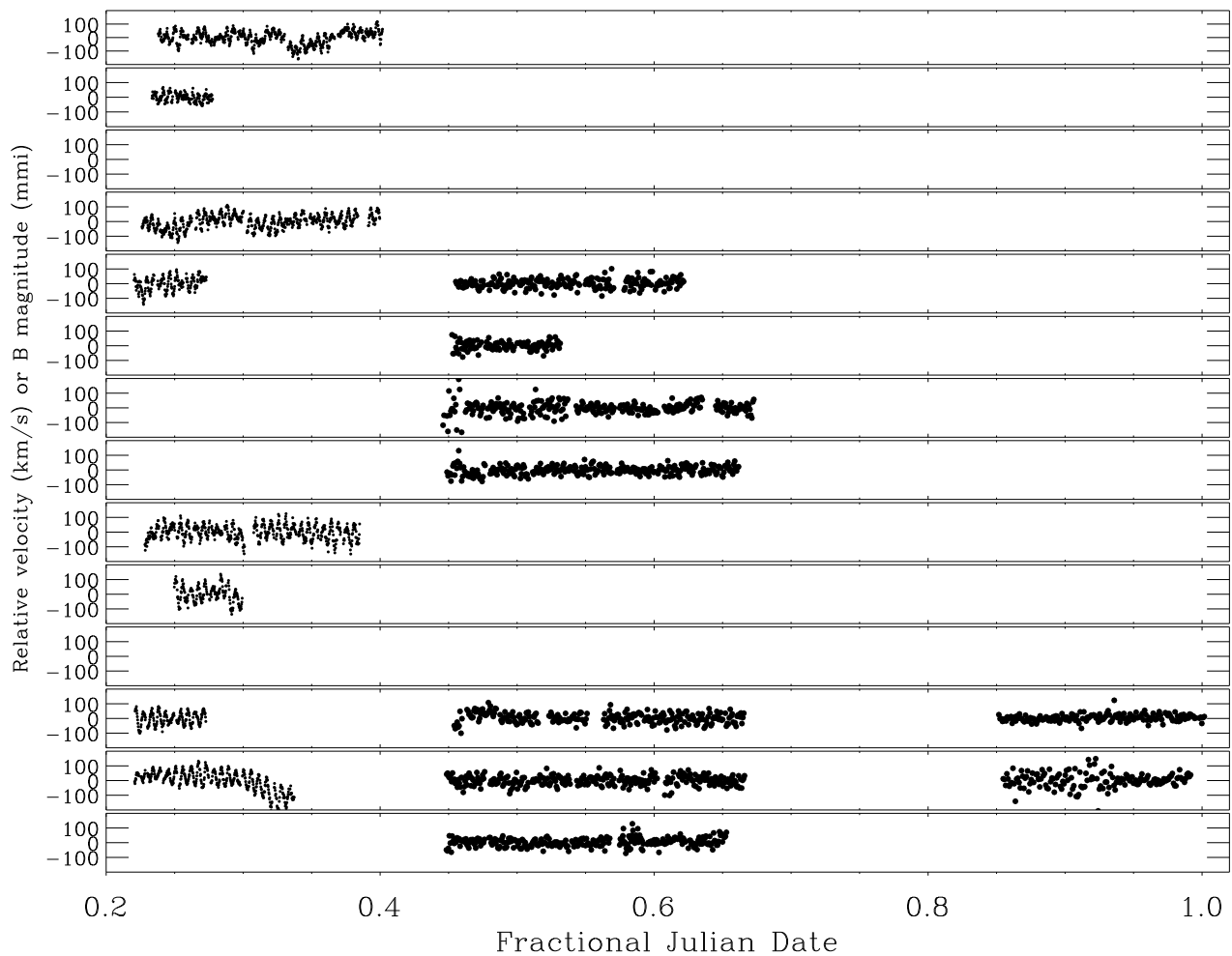}
\caption{Combined raw H$\beta$ velocity  and Johnson $B$ light curve
  for PG1605+072. Velocity points (in km\,s$^{-1}$) are large dots
  while photometry points are small dots. Photometry in this
  Letter is presented in 'millimodulation units' or mmi, where the
  light curve is normalized by its own mean level (see
  \citealt{ECpaperX}). Observations were made on JD 2451\,380-393
  (see Tables \ref{spec} and \ref{phot}). Note that positive velocity
  is redshifted by definition.}

\label{rawvel}
\end{figure}

\placefigure{rawvel}
\notetoeditor{rawvelocity.ps should be over two columns}

The photometric data were reduced by subtracting the sky background,
correcting for atmospheric extinction and normalising to the mean
intensity of the run, following a method similar to
\citet{ECpaperIV,ECpaperVI}. Slow drifts were not corrected
for in the photometry.

The H$\beta$ velocity curves and B magnitude light curves for each
night of observations are shown in Figure \ref{rawvel}. Oscillations
are not as evident in our velocity curve as they are in the light
curve, since the signal-to-noise per measurement is lower.

As is common with time-series measurements, the quality of the data
varies considerably through the data set. We therefore performed
frequency analysis using a weighted Fourier Transform, where weights
were assigned according to the local rms scatter \citep{K+F92}.

\section{Results and Discussion}

\begin{figure}[h]
\plotone{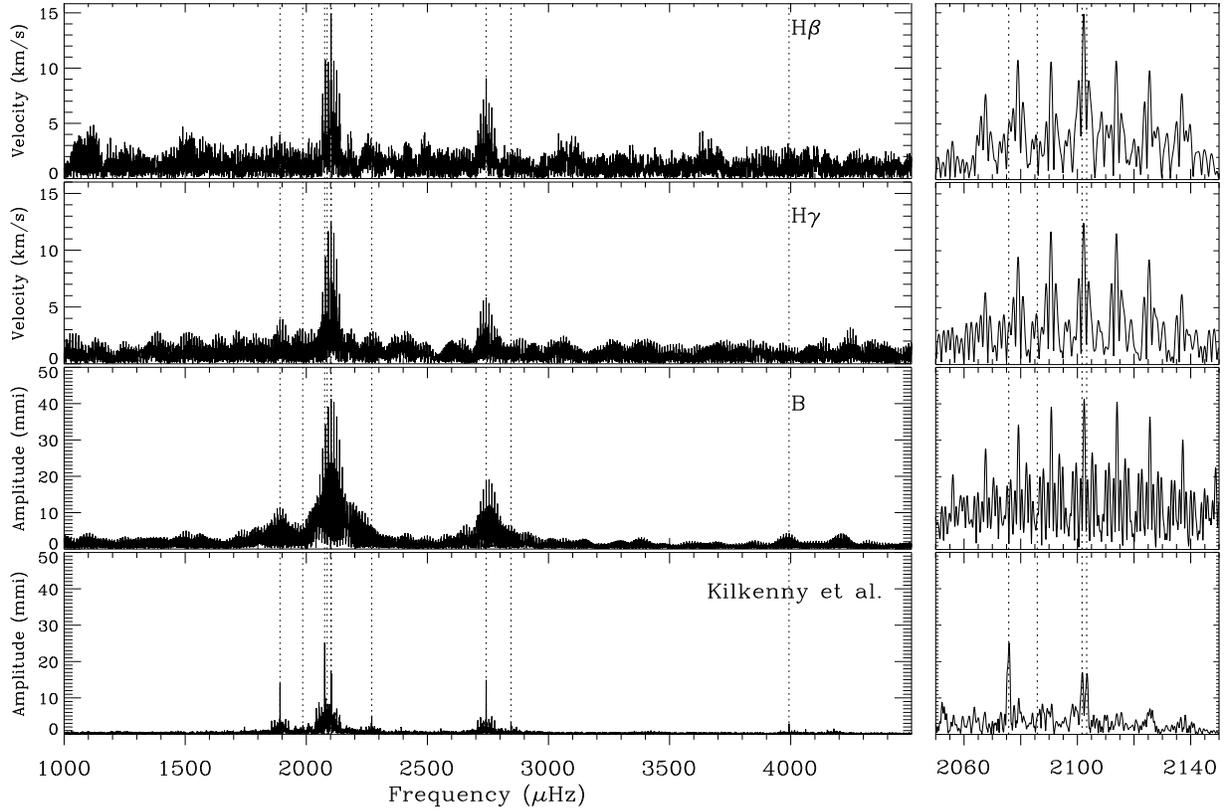}
\caption{Amplitude spectrum of PG1605+072 in the 1000--5000\,$\mu$Hz
  range. Panels on the left from top to bottom: H$\beta$ velocity;
  H$\gamma$ velocity (with no Stromlo data); $B$ band photometry;
  Photometry by \protect\citet{ECpaperX} in May, 1997 -- the dotted lines
  represent the 10 strongest modes found. Panels on the right show
  greater detail centred around 2100\,$\mu$Hz.}
\label{comp}
\end{figure}

The amplitude spectra of H$\beta$ and H$\gamma$ are shown in the top
two panels on the left of Figure \ref{comp}. The white noise level is
1.07\,km\,s$^{-1}$ for H$\beta$ and 0.93\,km\,s$^{-1}$ for H$\gamma$.
Oscillations are clearly visible at 2102\,$\mu$Hz and 2743\,$\mu$Hz
(2742.72\,$\mu$Hz in \citealt{ECpaperX}) in both lines but at 
1891\,$\mu$Hz (1891.42\,$\mu$Hz) only in H$\gamma$. It is unclear whether
this last effect is due to the low amplitude of the mode in H$\beta$ due to
beating between closely spaced modes. The bottom two panels
show our $B$ magnitude and white light amplitude spectra (from a 15-day
multisite campaign by \citet{ECpaperX} in May, 1997; white light
observations were obtained using a CuSO$_4$ filter) respectively,
and are included for comparison. The three modes previously mentioned
are all present in the $B$ magnitude spectrum. We
believe that the highest amplitude peak (14\,km\,s$^{-1}$ for
H$\beta$) at 2102\,$\mu$Hz may be a combination of the
2101.65\,$\mu$Hz and 2103.28\,$\mu$Hz frequencies found by
\citeauthor{ECpaperX} -- a longer time-series is needed to clarify this.

\placefigure{comp}
\notetoeditor{compspec.ps should be over two columns}

The panels on the right of Figure \ref{comp} show the
2050$-$2150\,$\mu$Hz region in greater detail. The highest amplitude
mode found by \citeauthor{ECpaperX} at 2075.76\,$\mu$Hz is not present
in any of our observations. It is not clear whether this is due to
variation in amplitudes over the 2.2 years between observations, or to
beating between very closely spaced modes. 

The $B$ magnitude and white light amplitudes differ by around 40\%,
which is to be expected in such hot, blue stars, considering that the
effective wavelength of white light is redder than the $B$ band. If
oscillation phase is wavelength dependent, then this would also reduce
the amplitude in white light (which is a much broader ``band'' than $B$).

\subsection{Oscillation Amplitudes}

In this Letter we present a qualitative comparison of luminosity and
velocity amplitudes. All amplitudes were
measured by the height of their peak. A more detailed quantative study
will be the subject of a subsequent paper.

\citet{K+B95}
derived a relationship between luminosity amplitude at a given
wavelength and velocity amplitude for classical and solar-like
oscillations. Using their scaling law, which assumes a black body
spectrum and adiabatic oscillations, we find that the velocity
amplitudes in PG1605+072 are 2--3 times lower than
expected. However a black body and adiabatic assumption is far from
valid in such a hot star when one is considering amplitudes in
visual. In a pure adiabatic assumption there is a 180\degr\ phase
difference between temperature and stellar radius (with maximum
brightness coinciding with minimum radius). For the mode at
2.742\,mHz (which do not appear to be contaminated by closely spaced
modes), we find a phase difference of (-75$\pm$10)\degr\ implying
that maximum brightness is slightly after maximum radius. More data are
needed to clarify this result.

\begin{figure}[h]
\plotone{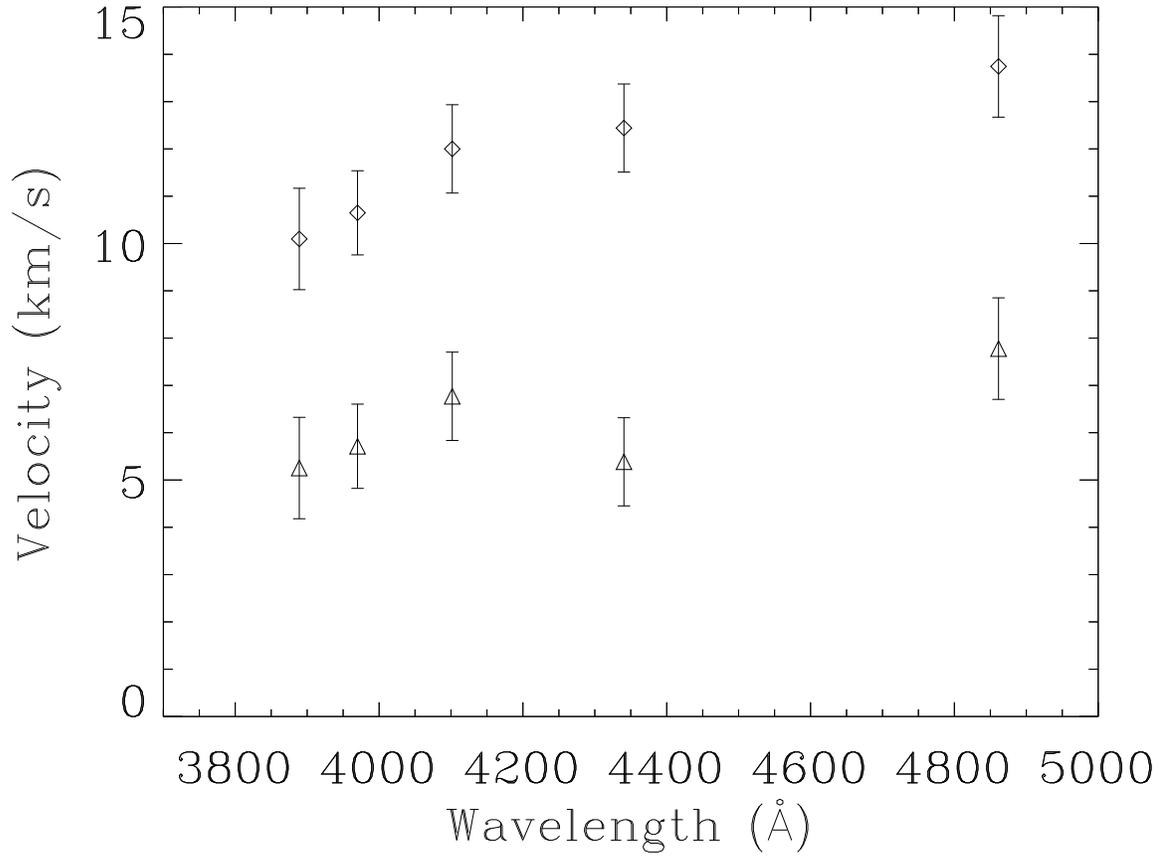}
\caption{Velocity amplitude vs wavelength for the 5 Balmer lines
  studied. The group of unresolved modes at 2.10 mHz is represented by
  diamonds, while the mode at 2.74 mHz is represented by
  triangles.}
\label{wave}
\end{figure}

\placefigure{wave}
\notetoeditor{wave-vs-amp.dat.ps should be over one column}

There is evidence to suggest
that the velocity amplitude is wavelength dependent. Figure \ref{wave} shows
that the amplitude of oscillations at two frequencies decreases
blueward of H$\beta$. The amplitude of H8 is almost 25\% smaller than
that of H$\beta$. If confirmed, this would be puzzling, given that all
the Balmer lines are formed in the same part of the atmosphere.

\section{Conclusions}

The results presented here show that times-series spectroscopy
of sdBs can be done using a medium-size (1.5\,m) telescope.
We have detected oscillations in velocity at previously published
photometric frequencies. 
A longer times-series would allow most of the closely spaced modes in
PG1605+072 to be resolved and thus help to
confirm the variation of velocity amplitude with wavelength. Further
studies may also determine whether the amplitudes of modes in this
star are variable over time, or whether the apparent shift in primary
pulsation mode is due to beating. We will also examine the behaviour
of equivalent width and other line profile variations.

\acknowledgements

This work was supported by an
Australian Postgraduate Award (SJOT),
the Australian Research Council, the Danish National Science
Research Council through its Center for Ground-based Observational
Astronomy, and the Danish National
Research Foundation through its establishment of the Theoretical
Astrophysics Center.

\newpage
\begin{table}[h]
\caption{Spectroscopic observations of PG1605+072. LS = La Silla; MS =
  Mt Stromlo.}
\label{spec}
\vspace{0.2cm}
\begin{center}
\begin{tabular}{lccc}
\hline
UT-date & Site & No. of & No. of \\
& & hours & spectra \\
\hline

1999 July 23--4 & LS & 4.01 & 210 \\
1999 July 24--5 & LS & 1.91 & 105 \\
1999 July 25--6 & LS & 5.46 & 280 \\
1999 July 26--7 & LS & 5.12 & 280 \\
1999 July 30--1 & LS & 5.09 & 260 \\
1999 July 31 & MS & 3.61 & 168 \\
1999 July 31--1 & LS & 5.23 & 280 \\
1999 Aug 01 & MS & 3.30 & 155 \\
1999 Aug 01--2 & LS & 4.92 & 280 \\
\hline
\textbf{total} & & 38.65 & 2018 \\

\end{tabular}
\end{center}

\end{table}

\newpage
\begin{table}[h]
\caption{Photometric observations of PG1605+072, taken at the Sutherland site of the SAAO.}
\label{phot}
\vspace{0.2cm}
\begin{center}
\begin{tabular}{lcccc}
\hline
UT-date &No. of & No. of \\
 & hours & obs. \\
\hline
1999 July 20 & 3.93 & 666 \\
1999 July 21 & 1.07 & 182 \\
1999 July 23 & 4.17 & 679 \\
1999 July 24 & 1.27 & 211 \\
1999 July 28 & 3.77 & 597 \\
1999 July 29 & 1.19 & 199 \\
1999 July 31 & 1.24 & 210 \\
1999 Aug 01 & 2.80 & 454\\
\hline
\textbf{total} & 19.44 & 3198 \\

\end{tabular}
\end{center}
\end{table}

\newpage

\newpage

\newpage

\end{document}